# Automatic Multi-Class Cardiovascular Magnetic Resonance Image Quality Assessment using Unsupervised Domain Adaptation in Spatial and Frequency Domains

Shahabedin Nabavi[1], Hossein Simchi[1], Mohsen Ebrahimi Moghaddam[1], Alejandro F. Frangi[2], Ahmad Ali Abin[1]

1- Faculty of Computer Science and Engineering, Shahid Beheshti University, Tehran, Iran.
2- Centre for Computational Imaging and Simulation Technologies in Biomedicine (CISTIB), School of Computing, University of Leeds, UK.

**Abstract**

Population imaging studies rely upon good quality medical imagery before downstream image quantification. This study provides an automated approach to assess image quality from cardiovascular magnetic resonance (CMR) imaging at scale. We identify four common CMR imaging artefacts, including respiratory motion, cardiac motion, Gibbs ringing, and aliasing. The model can deal with images acquired in different views, including two, three, and four-chamber long-axis and short-axis cine CMR images. Two deep learning-based models in spatial and frequency domains are proposed. Besides recognising these artefacts, the proposed models are suitable to the common challenges of not having access to data labels. An unsupervised domain adaptation method and a Fourier-based convolutional neural network are proposed to overcome these challenges. We show that the proposed models reliably allow for CMR image quality assessment. The accuracies obtained for the spatial model in supervised and weakly supervised learning are $99.41 \pm 0.24$ and $96.37 \pm 0.66$ for the UK Biobank dataset, respectively. Using unsupervised domain adaptation can somewhat overcome the challenge of not having access to the data labels. The maximum achieved domain gap coverage in unsupervised domain adaptation is $16.86\%$. Domain adaptation can significantly improve a 5-class classification task and deal with considerable domain shift without data labels. Increasing the speed of training and testing can be achieved with the proposed model in the frequency domain. The frequency-domain model can achieve the same accuracy yet $1.548$ times faster than the spatial model. This model can also be used directly on k-space data, and there is no need for image reconstruction.

## 1- Introduction

Cardiovascular magnetic resonance (CMR) imaging has many clinical applications as a powerful non-invasive diagnostic tool. This imaging modality can assess cardiac function and support the diagnosis of cardiovascular diseases, including coronary artery disease, cardiomyopathy, congenital and valvular disease (Bandettini and Arai 2008). CMR imaging is the gold standard in many diagnostic and therapeutic applications (Kumar, Patton et al. 2010, Salerno, Sharif et al. 2017). In addition, CMR is used in most cardiovascular population imaging studies, given its optimal trade-off non-invasiveness and accuracy. Routine quantitative analysis of CMR is hindered, amongst other factors, by imaging artefacts. These artefacts affect the cardiac image analysis, rendering cardiac anatomy or functional parameters inaccurate for diagnostic or epidemiological studies. The increasingly wider availability of CMR expertise in mainstream cardiology and the development of novel MRI protocols have reduced the incidence of these artefacts yet their presence remains unavoidable and visual determination of image quality is impractical in busy cardiovascular imaging departments. Hence, automated approaches to detect and identify these artefacts remains highly desirable.

Many artefacts may appear in CMR images due to patient's movement during imaging or internal organ motions, limitations in imaging systems, suboptimal acquisition settings, etc. Common reasons for CMR artefacts are respiratory and cardiac motion, blood flow, Gibbs ringing effect, aliasing, chemical shift, and field inhomogeneities (Ferreira, Gatehouse et al. 2013). After each image acquisition and over an extended period, image quality needs to be assessed to ensure optimal diagnostic value and minimal image quality drifts over time. Human quality assessment of CMR images can be time-consuming and costly. The many images acquired during an imaging examination require automatic machine learning methods for image quality assessment. Therefore, we propose an automated CMR image quality assessment method to detect different artefacts in this study. Besides image quality control, it also reveals the type of artefact. Thus, if the imaging needs to be repeated, it makes the imaging technicians aware of the source of the problem.

The contributions of this study are:

(1) An automated image quality assessment approach is proposed that handles four common types of CMR artefacts. Respiratory motion, cardiac motion, Gibbs ringing, and aliasing artefacts are examined. In this approach, CMR images with different views, including short-axis cine CMR, two, three, and four-chamber long-axis images, are considered input. The input of the proposed method can also be k-space to check the quality with no image reconstruction. Therefore, the proposed method can be used both in spatial and frequency domains.

(2) The approach is evaluated on several datasets to avoid biasing the assessment and understand the generalisability of the process. Since access to data labels is usually limited, a method is proposed to generate a training set by modelling imaging artefacts using k-space manipulation, producing realistic corrupted images. A deep domain adaptation method in spatial and frequency domains is then presented and evaluated for unavailable data labels in some datasets. These proposed methods are assessed in supervised, weakly supervised, and unsupervised manners.

The remainder of this article is organised as follows. Section II summarises previous related studies. Section III introduces proposed methods for adding artefacts to CMR images using k-space manipulation and presents deep learning models for CMR image quality assessment. The details and how to perform the experiments are described in Section IV. The results of the evaluation and validation of the proposed models are given in Section V. Finally, the results are discussed in Section VI to explain the essential characteristics of the proposed models before concluding in Section VII.

## 2- Related works

Objective image quality assessment methods are divided into complete reference, reduced-reference, and no-reference according to the availability of the reference image. Availability of the reference image means that the quality of each received image is compared to the reference image. The main challenge in the quality assessment of medical images is the impossibility of accessing this reference image, so estimating the quality of medical images is generally established without reference. Chow and Paramesran reviewed the methods of assessing the quality of medical images and specifically predicted that the next generation of methods would be no-reference (Chow and Paramesran 2016).

Assessing the quality of medical images plays an essential role in the design and fabrication of imaging devices. It plays a crucial role in the diagnosis and treatment planning for patients. In the last decade, some popular image quality assessment methods, such as peak signal to noise ratio (PSNR) and structural similarity index measure (SSIM) metrics, have been used to evaluate the quality of medical images (Ding 2018). Unlike classic distortions in natural images, distortions in medical images are dramatically different, making it challenging to use existing methods to evaluate the quality of natural images for medical applications. In medical imaging, image quality is related to various factors such as contrast, resolution, noise, and various image artefacts. The medical image quality can be related to the imaging method, the dose of radiation used for imaging, the area to be imaged or field of view (FOV), the characteristics of the imaging device, and the imaging setups specified by the technicians. In addition, there are different contents in medical than natural images, which also vary depending on the imaging modality. Considering these issues, exclusive methods appropriate to each modality and area to be imaged should be proposed for medical images. Also, the need to provide medical images with the highest possible quality to optimise diagnostic decisions has been accepted. Therefore, the medical image quality assessment is multifactorial. The large volume of medical images acquired for a patient also highlights the need to use automated methods in quality assessment to avoid the additional costs associated with physicians evaluating images (Zhang, Cavaro-Ménard et al. 2015).

Several studies have examined the criteria for evaluating the quality of natural images and their correlation with the radiologists' viewpoint in magnetic resonance imaging. These studies merely examine the applicability of these evaluation criteria in magnetic resonance imaging and do not specifically focus on specific artefacts (Chow, Rajagopal et al. 2016, Chow and Rajagopal 2017, Mason, Rioux et al. 2019). One of the most challenging artefacts highly regarded by researchers in both detection and correction is motion artefact. This artefact can occur for various reasons, including patient displacement during imaging, internal organ motions due to breathing, heart movements, and blood flow (Zaitsev, Maclaren et al. 2015). In a study by Kustner et al. (Küstner, Liebgott et al. 2018), the head and abdomen MR images were examined in 16 volunteers to automatically detect motion artefacts. For this purpose, a convolutional neural network (CNN) architecture and the input of overlapping patches of the original images have been used. The accuracy for detecting motion artefacts in the head area is 97% and in the abdomen area is 75%. In a study (Lorch, Vaillant et al. 2017) that detects motion artefact in MR images of the head and chest area using decision trees and the generation of artificially degraded images by manipulating the k-space, the accuracy of the classification is between 75 to 100% depending on the extracted features, the images, and other parameters. In the study of Graham et al. (Graham, Drobnjak et al. 2018), diffusion-weighted MRI was used to detect intra-volume movement artefacts. This study detects the artefacts using CNNs by considering a small volume of labelled images artificially augmented as the dataset.

In the study (Obuchowicz, Oszust et al. 2020), a no-reference quality assessment criterion is proposed for MR images based on entropy analysis, evaluated on 70 MR images from different body parts. The results have been compared with ten advanced methods in this field. The mean square error (MSE) of this method is 0.53. In another study (Esteban, Birman et al. 2017), an automated process for detecting the quality of MR images of the brain based on the random forest method has been proposed. The accuracy for the data set consisting of 1367 images has been reported to be about 76%. In a study by Samani et al. (Samani, Alappatt et al. 2020), a deep learning method using CNNs has been proposed that examined several artefacts simultaneously in a data set consisting of 332,000 slices of diffusion-weighted MRI. The accuracy is 98% for this proposed method.

The methods that precisely assess the quality of CMR images are discussed. In a study by Tarroni et al. (Tarroni, Oktay et al. 2018), an automated process for determining the quality of short and long-axis CMR images is proposed, examining image quality related to full heart coverage in the images, motion artefact, and contrast estimation. This study has been performed on two datasets, one with 3000 samples for training and testing processes and 100 samples for testing based on random forests. This study shows the sensitivity and specificity of 88 and 99% for the diagnosis of full cardiac coverage and 85 and 99% for the diagnosis of motion artefact. In another similar study, the same method has been evaluated with a larger dataset (Tarroni, Bai et al. 2020). In the study (Oksuz, Ruijsink et al. 2019), motion artefacts due to breathing and cardiac movements were investigated. In this study, k-space manipulation has been used to increase the volume of distorted data, and a CNN architecture has been used to learn the automatic detection model. The results on 3510 CMR images show an area under the ROC curve (AUC) of 0.89. In a study by Zhang et al. (Zhang, Gooya et al. 2018), the problem of full left ventricular coverage using CNNs has been investigated in a data set of over 5,000 cases, with an error rate of less than 5%. In this study, short-axis cine CMR images have been received as input. Two parallel CNN architectures have been used to determine the presence or absence of images related to the apex and the basal of the heart.

The main innovation of this study is using a particular layer in the end part of the proposed architecture to improve the classification of images, which has been introduced as a Fisher discriminative layer. This layer tries to minimise within-class variance and maximise the distance of between-class means. In another study (Zhang, Pereañez et al. 2018), the use of an adversarial learning approach based on CNNs to detect the entire left and right ventricular coverage of the heart has been investigated. This method has been studied on three datasets, and the results show the superiority of this method over previous approaches. In (Zhang, Gooya et al. 2017), generative adversarial networks have been evaluated to detect full left ventricular coverage in a dataset consisting of over 6,000 samples. The accuracy obtained in this study has been reported to be about 90%. In a study by Osadebey et al. (Osadebey, Pedersen et al. 2018), image quality assessment in the brain and cardiovascular images has been investigated. In this study, 16 volumes of CMR images have been used, and the method is based on the handmade feature extraction of four types of additive noise. A summary of these related studies is given in Table 1.

**Table 1. Summary of related studies in CMR image quality assessment.**

| Reference | Modality | | Proposed Method | Type of Artefact | | | | # of Samples | Dataset | Evaluations |
|---|---|---|---|---|---|---|---|---|---|---|
| | Short Axis | Long Axis | | Heart Coverage | Motion | Contrast Estimation | Others | | | |
| (Tarroni, Oktay et al. 2018) | ✓ | ✓ | Decision Forest | ✓ | ✓ | ✓ | | Over 3,000 cases | -UK Biobank<br>-UK Digital Heart Project | Sensitivity= 88%<br>Specificity= 99% |
| (Tarroni, Bai et al. 2020) | ✓ | ✓ | Decision Forest | ✓ | ✓ | ✓ | | 19,265 samples | -UK Biobank | - |
| (Oksuz, Ruijsink et al. 2019) | ✓ | | CNN and LSTM | | ✓ | | | 3,510 samples | -UK Biobank | AUC=0.89 |
| (Zhang, Gooya et al. 2018) | ✓ | | CNN | ✓ | | | | Over 5,000 samples | -UK Biobank | Loss= less than 5% |
| (Zhang, Pereañez et al. 2018) | ✓ | ✓ | CNN and GAN | ✓ | | | | Over 5,000 samples | -UK Biobank<br>- DETERMINE<br>- MESA | Accuracy=92% |
| (Zhang, Gooya et al. 2017) | ✓ | | GAN | ✓ | | | | Over 6,000 samples | -UK Biobank | Accuracy=92% |
| (Osadebey, Pedersen et al. 2018) | ✓ | | Handmade Feature Extraction | | | | ✓ | 16 volumes | - | - |

## 3- Method

This section describes methods for adding artefacts to images using k-space manipulation. It presents deep learning-based models for identifying CMR image artefacts, including respiratory and cardiac motion, aliasing and Gibbs ringing artefacts. Initially, methods for adding artefacts to CMR images were proposed to access data labels due to the challenges of human data labelling. These methods are based on the k-space manipulation to make realistic distorted images. Then, supervised, weakly supervised and unsupervised training models are described. These models are proposed in both spatial and frequency domains, and details of all models are described. For conditions where the data labels are not accessible, an unsupervised domain adaptation method is presented.

### 3-1- Artefacted CMR image ground truth via k-space manipulation

Due to hardware and software restrictions, CMR images may be associated with artefacts. These artefacts can seriously affect image quality and thus the ability to make a proper diagnosis. According to the published statistics, motion and Gibbs artefacts are the most common (Budrys, Veikutis et al. 2018). Therefore, based on the fact that respiratory and cardiac motion artefacts, Gibbs ringing and aliasing are the most common artefacts in CMR images, these four types of artefacts are examined in this study (Ferreira, Gatehouse et al. 2013).

Subjectively determining the artefact in the CMR images is a time-consuming and laborious task that requires several human observers. Since this study needs corrupted images, we used the k-space degradation methods to generate synthetic but realistic images. Therefore, long-axis and short-axis cine CMR images acquired using Cartesian sampling were used to add respiratory and cardiac motion, aliasing and Gibbs ringing artefacts. These sections explain how to corrupt images using k-space manipulation.

#### 3-1-1- Respiratory motion artefacts

Similar to previous studies (Lorch, Vaillant et al. 2017, Oksuz, Ruijsink et al. 2019), a sinusoidal pattern translation is first created in the reference image, and the translated image is stored. Then, with the Fourier transform, the reference and translated images are transferred to k-space. The lines of k-spaces corresponding to the reference and the translated images are combined in this step based on a sine pattern. To generate images with different severities of this artefact, the translation and the sinusoidal pattern of combinations of k-space lines are considered randomly. Finally, the combined k-space is reconstructed into the final image to generate the degraded image with this artefact. Figure 1 shows how to develop a corrupted slice with this method.

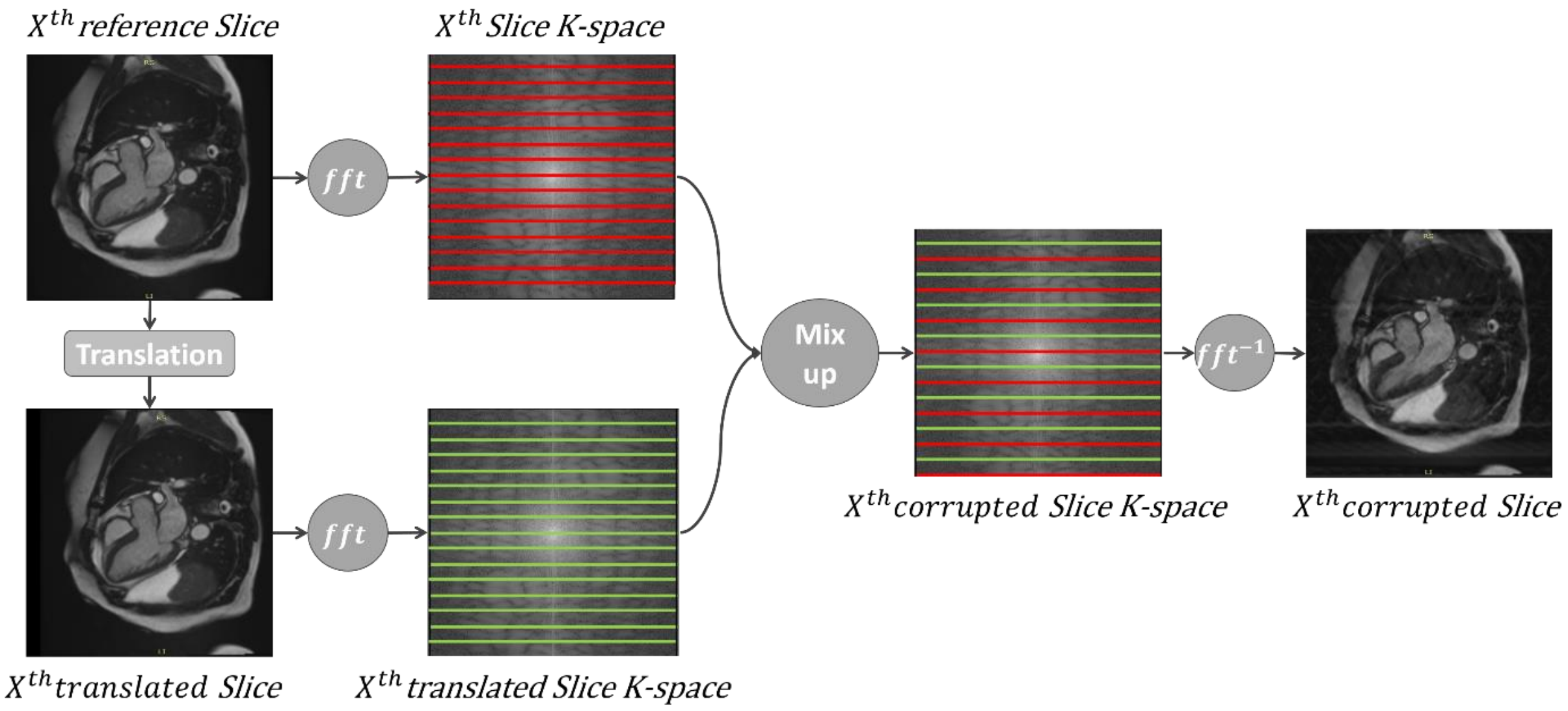

**Figure 1. How to corrupt a slice using k-space manipulation for generating respiratory artefact. A one-dimensional translation occurs in the reference slice. Then the k-spaces corresponding to the reference and the translated slices are combined with the sine pattern to produce the degraded image.**

### 3-1-2- Cardiac motion artefacts

Similar to (Oksuz, Ruijsink et al. 2019), short-axis cine CMR images need to be used to generate degraded images with cardiac motion artefacts. For this purpose, a temporal sequence is separated from the short-axis cine images that fully cover the heart from the basal to the apex. The images in this sequence are then transferred to the corresponding k-space using the Fourier transform. By replacing the k-space lines of the Cartesian sampled slices in this sequence, degraded images can be generated with this artefact. Synthetic images generated by this method are realistic because, during image acquisition, a similar pattern occurs due to mis-triggering in the lines of temporal k-spaces. Several lines were replaced randomly in consecutive k-spaces to simulate various levels of artefact severity. Figure 2 shows the process of image corruption with this artefact.

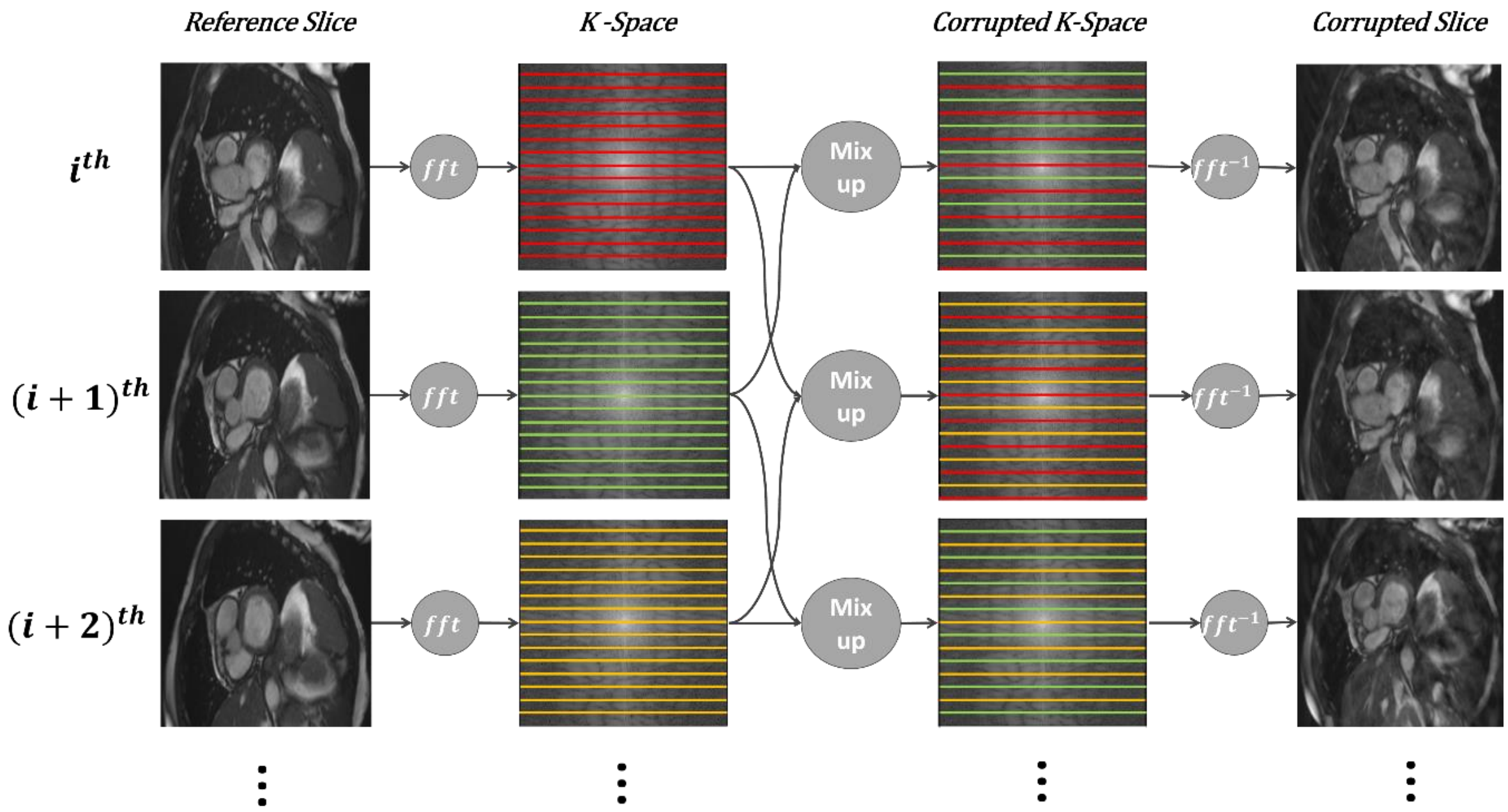

**Figure 2. An example of how to corrupt images with cardiac motion artefact using k-space manipulation. Replacing some of the k-space lines in a temporal sequence of short-axis cine CMR images is randomly done to generate degraded images with cardiac motion artefacts.**

### 3-1-3- Gibbs ringing artefacts

Gibbs ringing or truncation artefacts are due to using Fourier transforms in the reconstruction of the final image. Since to generate an image in magnetic resonance imaging, we have to take a limited sample of the generated signals, so when reconstructing the final image, we have to approximate it using a relatively small number of harmonics in the Fourier representation of the image. Therefore, truncated Fourier series approximation of a discontinuous signal shows overshoot and ringing near the discontinuities of the primary signal (Czervionke, Czervionke et al. 1988). This case is called the Gibbs phenomenon (Gibbs 1899), which causes parallel lines and curves in the final image. Image filtering can be used with an ideal low-pass filter in the frequency domain to add this artefact with different severities to the final image (Bovik and Acton 2009). Thus, the image is transferred to k-space by Fourier transform, and then k-space is filtered by an ideal low-pass filter. The radius of this low-pass filter was considered randomly for different samples to achieve more variety of this artefact in the final images. The procedure for the k-space manipulation to generate corrupted images is shown in Figure 3.

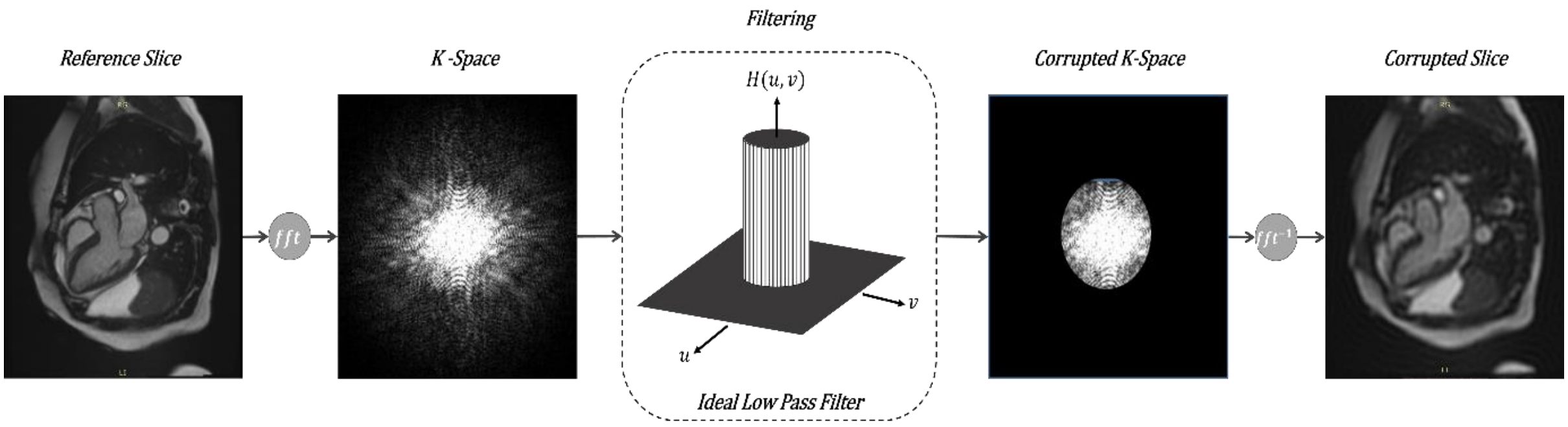

**Figure 3. How to manipulate k-space to add Gibbs ringing artefacts to images. The corresponding k-space can be filtered with an ideal low-pass filter to add this artefact to the image.**

### 3-1-4- Aliasing artefacts

Aliasing or wrap-around artefact occurs when the field of view (FOV) is smaller than the area being imaged. Here, the parts outside the FOV are reflected on the other side of the image. It is enough to undersample the k-space to add this artefact to the images by manipulating the k-space. Because the k-space of the images used is Cartesian sampled, the corrupted image can be obtained by alternately deleting rows or columns of k-space corresponding to an image (Huang, Seethamraju et al. 2015). An example of image degradation in this way is shown in Figure 4.

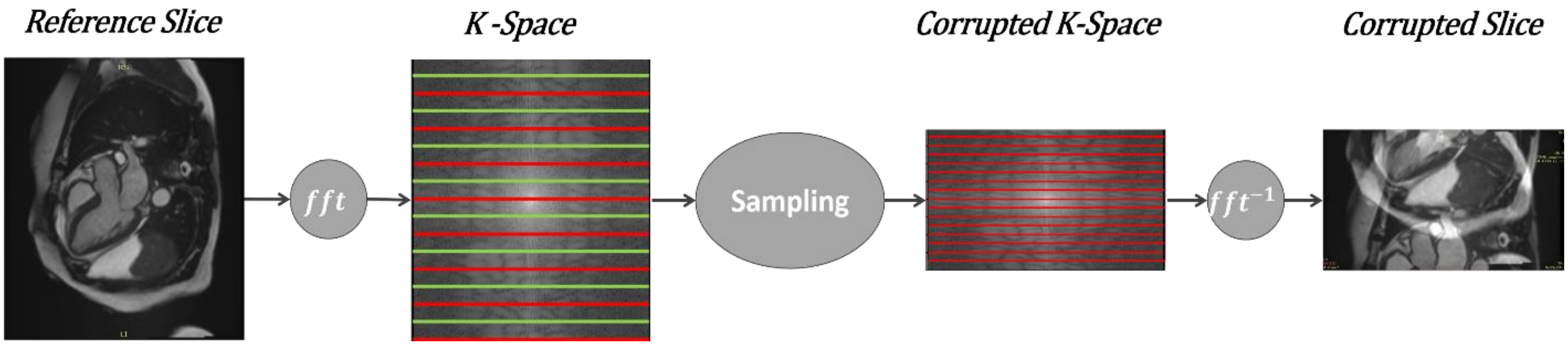

**Figure 4. How to add aliasing artefact to CMR images using k-space manipulation. A corrupted image with the aliasing artefact can be obtained by alternately deleting lines of the k-space.**

## 3-2- Deep learning model for automated CMR image quality assessment

In this Section, we propose a deep learning model for automatic CMR image quality assessment. First, this model is proposed to identify the artefacts in the spatial domain when data labels are available. The model details for domain adaptation are presented when we use the trained model to test a new dataset. Finally, the proposed model in the frequency domain is described to directly use the model on the raw data of k-space before image reconstruction.

The proposed model evaluates the quality of each of the 2D slices of a 3D volume of CMR images. The reason for presenting the 2D model is that the artefacts studied in this research have features that can be seen in each 2D slice. Although some previous studies have identified motion artefacts with 3D models like (Oksuz, Ruijsink et al. 2019), the diversity of artefacts in the current research and the need for a general model to distinguish between types of artefacts led us to use the 2D model.

### 3-2-1- Spatial domain

We use a deep learning model to identify artefacts in the spatial domain. This model generally has three essential parts: (1) Feature extraction from CMR images (2) Classification of the image artefacts (3) Domain label predictor for domain adaptation. Figure 5 shows the proposed model in the spatial domain.

The feature extraction section receives the input images with various views, including two, three and four-chamber long-axis and short-axis views. The dimensions of the input images are resized to $90 \times 90$ to control the computational overhead. Four 2D convolutional layers with the details specified in Figure 5 are considered for image feature extraction. Each pair of convolutional layers is followed by a max-pooling layer and a dropout layer with a probability of $0.25$. Finally, a flatten followed by a dense layer are at the end of this Section, which presents the extracted features in a $512$ vector.

The feature vector generated by the feature extractor part is received as input in the label classification Section. This Section follows a dropout layer with a probability of $0.5$ by two dense layers responsible for identifying the artefact labels. Five labels related to respiratory and cardiac motions, aliasing, Gibbs ringing and without artefacts are determined by five terminal neurons. If training data labels are available and supervised or weakly supervised learning is on the agenda, we will only use the feature extraction and label classification sections. However, we need to use domain adaptation.

Domain adaptation refers to the concept of learning a discriminative classifier when there is a shift in distributing training and testing data distribution. Domain adaptation approaches are classified into two categories, unsupervised and semi-supervised, based on the possibility of accessing the data labels of the target dataset. The former is when the target dataset is unlabelled, and the latter is when a few samples are labelled. These approaches allow the mapping between the source and target domains to be learned when the target domain is unlabelled or partially labelled. It means that the classifier can be trained on a dataset as the source domain and then used to test the target domain (Kouw and Loog 2019). The main idea of domain adaptation is inspired by (Ganin and Lempitsky 2015) and customised for the current study. All three subnets required for feature extraction, artefact classification, and domain label prediction have been redesigned and implemented to meet the objectives of this study.

If the data labels are not available in a database and there is a domain shift between datasets, we should also use the domain classification Section for unsupervised domain adaptation. This Section follows a gradient reversal layer (GRL) by three dense layers with the details specified in Figure 5. The GRL layer acts as an identical transformation in forwarding propagation while multiplying the gradient received from the dense layers by a negative value in backpropagation. The reason for using this layer is there are conflicting goals between the sections. If there are two datasets, one as a source with access to the data labels and the target without data labels, the feature extraction Section must extract domain-invariant features between different datasets at training time. The source data must have a domain label and an artefact label.

In contrast, the target data have the domain label, which differs from the source domain label, to achieve these features. Therefore, the feature extraction Section maximises the domain classifier loss and minimises the artefact label classifier loss. Maximising the domain classifier loss makes the distribution of features obtained from the two domains more similar. Thus, after completing the model training, the model will extract features from the target dataset identical to the source set at testing time.

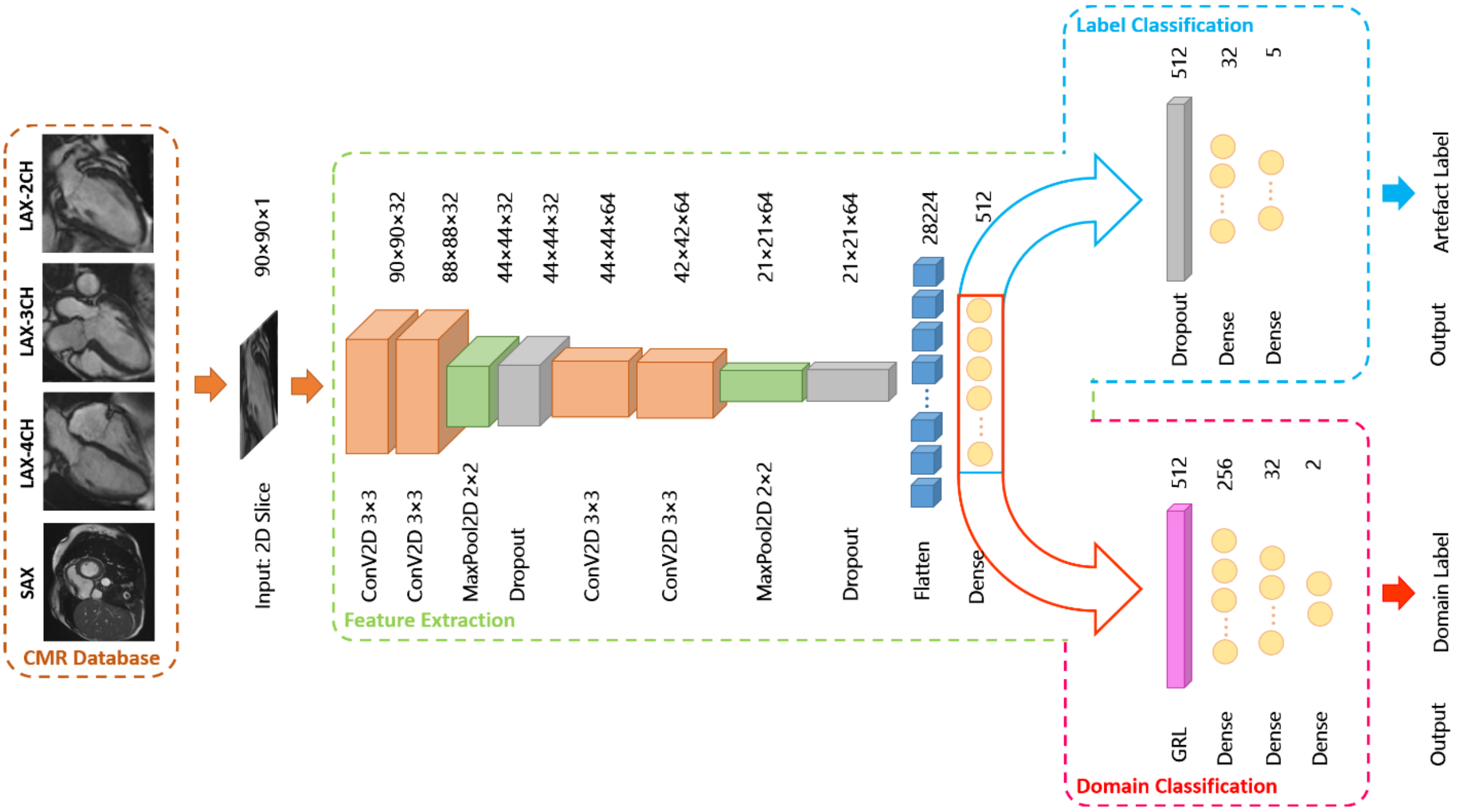

**Figure 5. The proposed deep learning model for the CMR image quality assessment in the spatial domain. The feature extraction section, which consists of 2D convolutional layers, extracts the features of images and finally provides them to the other parts in a 512 vector. The label classification Section specifies the artefact in the images. The domain classifier Section is also used to extract domain-invariant features in domain adaptation.**

### 3-2-2- Frequency domain

Two main concerns led us to provide a model in the frequency domain for CMR image quality assessment: (1) Possibility to use the proposed method in the frequency domain on the raw data of k-space before image formation (2) Increase the speed of training and testing operations when the model may be used for image quality assessment in large datasets such as UK Biobank. The proposed model is shown in Figure 6.

This proposed model uses Fourier-based CNNs (FCNNs) (Pratt, Williams et al. 2017). FCNNs can reduce computational costs and compensate for the slight decrease in accuracy by increasing training data. These networks can be used well when working on large datasets and analysing k-space data. Therefore, the convolution operation in the frequency domain can be changed to a pointwise multiplication in this model by considering Equation (1).

$$DFT\big(f(x,y) * h(x,y)\big) = DFT(f(x,y)) \cdot DFT(h(x,y)) \qquad (1)$$

Where $DFT$ represents the discrete Fourier transform, $f(x,y)$ represents the input image, and $h(x,y)$ indicated the filter.

The k-space corresponding to the input image is pointwise multiplied in three filters filled with random complex numbers for feature extraction. This operation is repeated twice, and the size of the filters is the same as the size of the k-space. The output of these steps is given to the max-pooling layer, and the dropout with a probability of 0.25, flatten, and dense layers follow this layer. Finally, a 512 vector is transferred to the label classification Section; this section's details are shown in Figure 6.

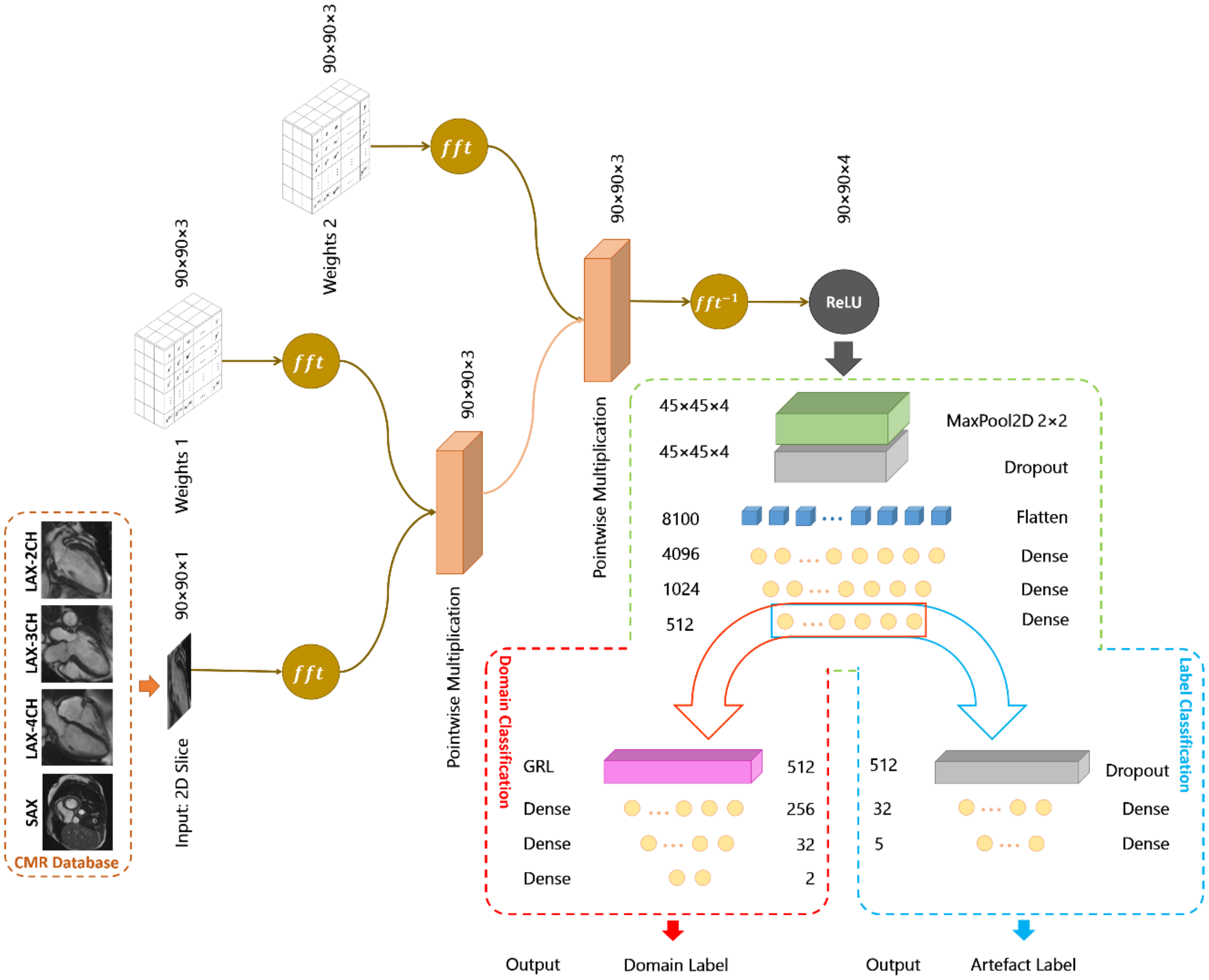

**Figure 6. The proposed deep learning model for the CMR image quality assessment in the frequency domain. Instead of using the convolution layer, this model uses pointwise multiplication, which dramatically reduces the complexity of the feature extraction. A vector of 512 extracted features is then given to the label classification Section to identify the artefact label.**

## 4- Experimental setup

### 4-1- Dataset description

The main database of the current study is the CMR image database from the UK Biobank. Imaging data from over 6,000 subjects of this database were used to generate a dataset including CMR images corrupted by synthetically but realistic artefacts. CMR images were acquired using a wide clinical bore $1.5T$ MR system (MAGNETOM Aera, Syngo Platform VD13A, Siemens Healthcare, Erlangen, Germany) with an 18 channel anterior body surface coil ($45\ mT/m$ and $200\ T/m/s$ gradient system). 2D cine balanced steady-state free precession (bSSFP) short-axis and long-axis images have in-plane spatial resolution $1.8 \times 1.8\ mm$, slice thickness $8\ mm$ and slice gap $2\ mm$. More information about the imaging protocol is available in (Petersen, Matthews et al. 2015).

Besides UK Biobank, data from three other datasets have been used. One of these datasets is the cardiac MRI dataset from York University (YU). A total of 7980 short and long-axis images have been acquired from 33 subjects. Imaging has been performed with a GE Genesis Signa MR scanner using the Fast imaging employing steady-state acquisition (FIESTA) scanning protocol. The subjects studied in this research are all under 18 years old. The spatial resolution of each slice is $256 \times 256$ pixels. In the short-axis view, the number of frames is 20, and in the long-axis view, the number of slices varies between 8 and 15. The slice gap is also between $6$ and $13\ mm$. Details related to this dataset can be found in (Andreopoulos and Tsotsos 2008). The Universidad Carlos III (UCIII) dataset (Abascal, Montesinos et al. 2014) contains short-axis cardiac cine MR images of four small animals is another dataset used in this study. A $7T$ Bruker Biospec $70/20$ scanner has been used to scan these self-gated rat cardiac cine sequences (IntraGateFLASH). The number of acquired frames is 8 with spatial resolution $192 \times 192$. slice thickness is $1.2\ mm$. Finally, another dataset including short and long-axis CMR images was collected from Tehran hospitals. Images of 21 unknown subjects have been gathered in this dataset. The imaging parameters are different between CMR image volumes in this dataset due to its multicentre nature. Patient information and images related to this database will remain confidential. These datasets have been selected due to availability and a significant domain shift compared to UK Biobank.

For data augmentation in YU, UCIII, and CMR-Tehran datasets with fewer data, horizontal and vertical flip, random rotation with padding and brightness changing methods have been used. The angle of rotation and the number of brightness changes were considered random to diversify the data further.

### 4-2- Model training

The training process of the existing models was performed in several steps according to different defined experiments. Adam optimiser (Kingma and Ba May 2015) and cross-entropy loss function (Goodfellow, Bengio et al. 2016) have been used to train the models in spatial and frequency domains. The learning rate was set to $1 \times 10^{-3}$. Also, the training process has been carried out in 25 epochs and with batch size 128. To validate the model training process and increase the reliability of the results, experiments were performed using 4-fold cross-validation. The best architecture suitable for the current study, layer placements and selection of hyperparameters was made based on several investigations. The hardware platform is a computer equipped with an Intel Core i-7-6700, 16GB of RAM and an NVIDIA GeForce GTX TITAN X GPU. After the article is published, all developed codes will be released on GitHub.

### 4-3- Evaluation design

We performed several experiments to evaluate the accuracy and capability of the proposed methods. In the first group of these experiments, the proposed deep learning models in spatial and frequency domains are learned in a supervised learning manner. This way, 75% of the data is a training set, and the rest is for testing. Given one of the most fundamental challenges, namely the lack of access to labelled data, the second group of experiments is assumed to be weakly supervised. In this group of experiments, 25% of the data is used for training and the rest for testing. This experiment makes the results comparable. The third set of experiments is performed to evaluate the proposed unsupervised domain adaptation model. UK Biobank data is first considered a source set in these experiments, and other data sets are considered a target. The same experiments are then repeated by considering the YU dataset as the source. In the last group of experiments, the difference between training and testing speed and the accuracy in the spatial and frequency domains are evaluated. These experiments examine the capability of the proposed models in the CMR image quality assessment.

Five metrics, including accuracy (ACC), precision (PR), recall (RE), F-measure, and area under the ROC curve (AUC), given in Equations 2 to 6, are used to evaluate the results of the proposed models quantitatively.

$$ACC = \frac{TP + TN}{TP + FP + TN + FN} \quad (2)$$

$$PR = \frac{TP}{TP + FP} \quad (3)$$

$$RE = \frac{TP}{TP + FN} \quad (4)$$

$$F - measure = \frac{2 \times PR \times RE}{PR + RE} \quad (5)$$

$$AUC = \int_0^1 \Pr[TP]\,(v)\,dv \quad (6)$$

where TP, FP, TN, and FN indicate true positive, false positive, true negative, and false negative. AUC denotes the overall success of an experiment where $\Pr[TP]$ is a function of $v = \Pr[FP]$.

## 5- Results

### 5-1- Supervised learning

In the first experiment, all datasets, including reference images and corrupted images generated by k-space manipulation methods, were used to train and test the models in a supervised manner. After the image corruption by k-space manipulating techniques and data augmentation, 255775, 144411, 23638 and 6080 CMR images were prepared for UK Biobank, YU, CMR-Tehran, and UCIII datasets, respectively. The results to assess the quality of CMR images in a supervised manner for respiratory and cardiac motions, Gibbs ringing and aliasing artefacts are given in Table 2. 75% of the data were used for training and 25% for testing in the first experiment.

**Table 2. Results for supervised evaluation of the proposed CMR image quality assessment models are based on accuracy, precision, recall, F-measure and AUC metrics in spatial and frequency domains. Each dataset has been used as input of the models by adding artefacts based on the proposed methods of k-space manipulation and after data augmentation with the mentioned methods. Respiratory and cardiac motions, Gibbs ringing and aliasing artefacts were added to corrupt the images. Results are based on 4-fold cross-validation ($\boldsymbol{Mean \pm std}$).**

| Dataset \ Metrics | | Accuracy | Precision | Recall | F-measure | AUC |
|---|---|---|---|---|---|---|
| **UK Biobank** | Spatial | $99.41 \pm 0.24$ | $99.41 \pm 0.24$ | $99.41 \pm 0.24$ | $99.41 \pm 0.24$ | $99.91 \pm 0.06$ |
| | Frequency | $87.46 \pm 0.61$ | $87.59 \pm 0.64$ | $87.31 \pm 0.57$ | $87.45 \pm 0.61$ | $98.64 \pm 0.19$ |
| **YU** | Spatial | $75.78 \pm 9.17$ | $76.88 \pm 9.27$ | $75.04 \pm 9.06$ | $75.95 \pm 9.15$ | $91.26 \pm 5.44$ |
| | Frequency | $63.76 \pm 12.35$ | $67.94 \pm 12.61$ | $58.23 \pm 13.02$ | $62.70 \pm 12.92$ | $88.47 \pm 7.71$ |
| **CMR-Tehran** | Spatial | $67.87 \pm 4.10$ | $68.46 \pm 3.97$ | $67.38 \pm 4.13$ | $67.91 \pm 4.05$ | $87.23 \pm 2.32$ |
| | Frequency | $58.48 \pm 0.41$ | $62.83 \pm 0.65$ | $48.37 \pm 3.78$ | $54.60 \pm 2.46$ | $88.63 \pm 0.59$ |
| **UCIII** | Spatial | $89.46 \pm 1.01$ | $89.61 \pm 1.20$ | $89.03 \pm 0.62$ | $89.32 \pm 0.87$ | $97.10 \pm 0.51$ |
| | Frequency | $80.25 \pm 0.94$ | $83.91 \pm 1.16$ | $76.34 \pm 2.00$ | $79.94 \pm 1.50$ | $96.83 \pm 0.34$ |

To make the results comparable, we want to look at a scenario in the following experiment with weak supervision. One of the main challenges of machine learning methods is accessing data labels to learn the model. Data labelling, mainly in medical image analysis, requires a time-consuming and laborious process. Therefore, the first experiment was repeated if only 25% of the data in each database was labelled. Thus, in this experiment, 25% of the data were used for training and the remaining 75% for testing. The results of 4-fold cross-validation are given in Table 3 for spatial and frequency domains.

**Table 3. The weakly supervised evaluation of the proposed CMR image quality assessment models is based on accuracy, precision, recall, F-measure and AUC metrics in spatial and frequency domains. Each dataset has been used as input of the models by adding artefacts based on the proposed methods of k-space manipulation and after data augmentation with the mentioned methods. Respiratory and cardiac motions, Gibbs ringing and aliasing artefacts were added to corrupt the images. Results are based on 4-fold cross-validation ($Mean \pm std$).**

| Dataset \ Metrics | | Accuracy | Precision | Recall | F-measure | AUC |
|---|---|---|---|---|---|---|
| UK Biobank | Spatial | $96.37 \pm 0.66$ | $96.37 \pm 0.66$ | $96.37 \pm 0.66$ | $96.37 \pm 0.66$ | $99.07 \pm 0.13$ |
| | Frequency | $79.78 \pm 1.56$ | $80.36 \pm 1.58$ | $79.23 \pm 1.52$ | $79.79 \pm 1.54$ | $96.45 \pm 0.34$ |
| YU | Spatial | $53.25 \pm 6.06$ | $53.82 \pm 6.11$ | $52.43 \pm 6.23$ | $53.11 \pm 6.17$ | $77.68 \pm 4.75$ |
| | Frequency | $45.66 \pm 4.42$ | $49.86 \pm 6.45$ | $37.54 \pm 4.28$ | $42.67 \pm 4.19$ | $77.61 \pm 5.68$ |
| CMR-Tehran | Spatial | $54.04 \pm 4.64$ | $54.30 \pm 4.92$ | $52.36 \pm 6.54$ | $53.30 \pm 5.78$ | $79.20 \pm 2.61$ |
| | Frequency | $45.96 \pm 2.47$ | $49.06 \pm 4.30$ | $36.63 \pm 1.93$ | $41.92 \pm 2.62$ | $81.54 \pm 1.81$ |
| UCIII | Spatial | $79.35 \pm 6.22$ | $79.75 \pm 5.92$ | $78.92 \pm 6.57$ | $79.33 \pm 6.24$ | $92.23 \pm 2.69$ |
| | Frequency | $59.40 \pm 4.65$ | $66.98 \pm 6.26$ | $50.94 \pm 3.33$ | $57.83 \pm 4.25$ | $86.26 \pm 3.44$ |

## 5-2- Unsupervised domain adaptation

All parts of the Figure 5 model are used to perform the experiments in this Section. To learn the model, any sample of the source set has both the domain and artefact label, but instances of the target set have only the domain label. The model will learn domain-invariant features for the samples of source and target sets in the training process. Therefore, it is expected that in the test phase, the model can classify unseen samples of the target set more accurately than when domain adaptation is not used. To compare the results, three modes of implementation are considered. The highest accuracy is achieved when the model can receive artefact labels related to the samples of the target set during training. The least accuracy also clearly occurs when the model gets no instances of the target set at the time of training. The trained model is evaluated on the samples of the target set. Therefore, it is expected that the domain adaptation method can cover the gap between these two modes. In the first experiment, we consider a scenario where the UK Biobank dataset is a source set, and the other datasets are considered a target. The results of the first experiment are shown in Table 4.

**Table 4. Results of the first unsupervised domain adaptation experiment based on accuracy metric. In this experiment, the source set is UK Biobank, and the target sets are YU, CMR-Tehran and UCIII. Three training modes have been used, including training only on source set (lower bound), training on target set (upper bound), and using the proposed domain adaptation method in training. The numbers in parentheses in the row corresponding to the results of the proposed method indicate the gap coverage between the upper and lower bands. The images related to the UK Biobank, YU, CMR-Tehran, and UCIII datasets are 255775, 106026, 17901 and 4784, respectively.**

| Training mode \ Target set | YU | CMR-Tehran | UCIII |
|---|---|---|---|
| Source only | 26.58 | 15.90 | 30.50 |
| Proposed method | 32.44 (+11.91%) | 24.51 (+16.57%) | 32.82 (+3.93%) |
| Train on target | 75.78 | 67.87 | 89.46 |

In the second experiment of this section, the results are presented in a situation that the source set is YU, and the target sets are the other datasets. The results are given in Table 5.

**Table 5. Results of the second unsupervised domain adaptation experiment based on accuracy metric. In this experiment, the source set is YU, and the target sets are UK Biobank, CMR-Tehran and UCIII. Three training modes have been used, including training only on source set (lower bound), training on target set (upper bound), and using the proposed domain adaptation method in training. The numbers in parentheses in the row corresponding to the results of the proposed method indicate the gap coverage between the upper and lower bands. The images related to the UK Biobank, YU, CMR-Tehran, and UCIII datasets are 255775, 106026, 17901 and 4784, respectively.**

| Target set / Training mode | UK Biobank | CMR-Tehran | UCIII |
|---|---|---|---|
| Source only | 28.62 | 50.53 | 31.80 |
| Proposed method | 30.17 (+2.19%) | 43.88 (-38.35 %) | 41.52 (+16.86%) |
| Train on target | 99.41 | 67.87 | 89.46 |

Graphs for comparing the results of this Section's first and second experiments are shown in Figure 7. Figure 8 shows how the proposed model works in domain adaptation. In this figure, 512 feature vectors from the feature extraction Section before and after domain adaptation are shown side by side for 20 input image samples. This output is generated by considering the YU dataset as the source set and the UCIII dataset as the target set. As shown in this figure, after domain adaptation, the feature values of the source and target sets have been brought closer to each other, confirming the proper performance of the domain adaptation method.

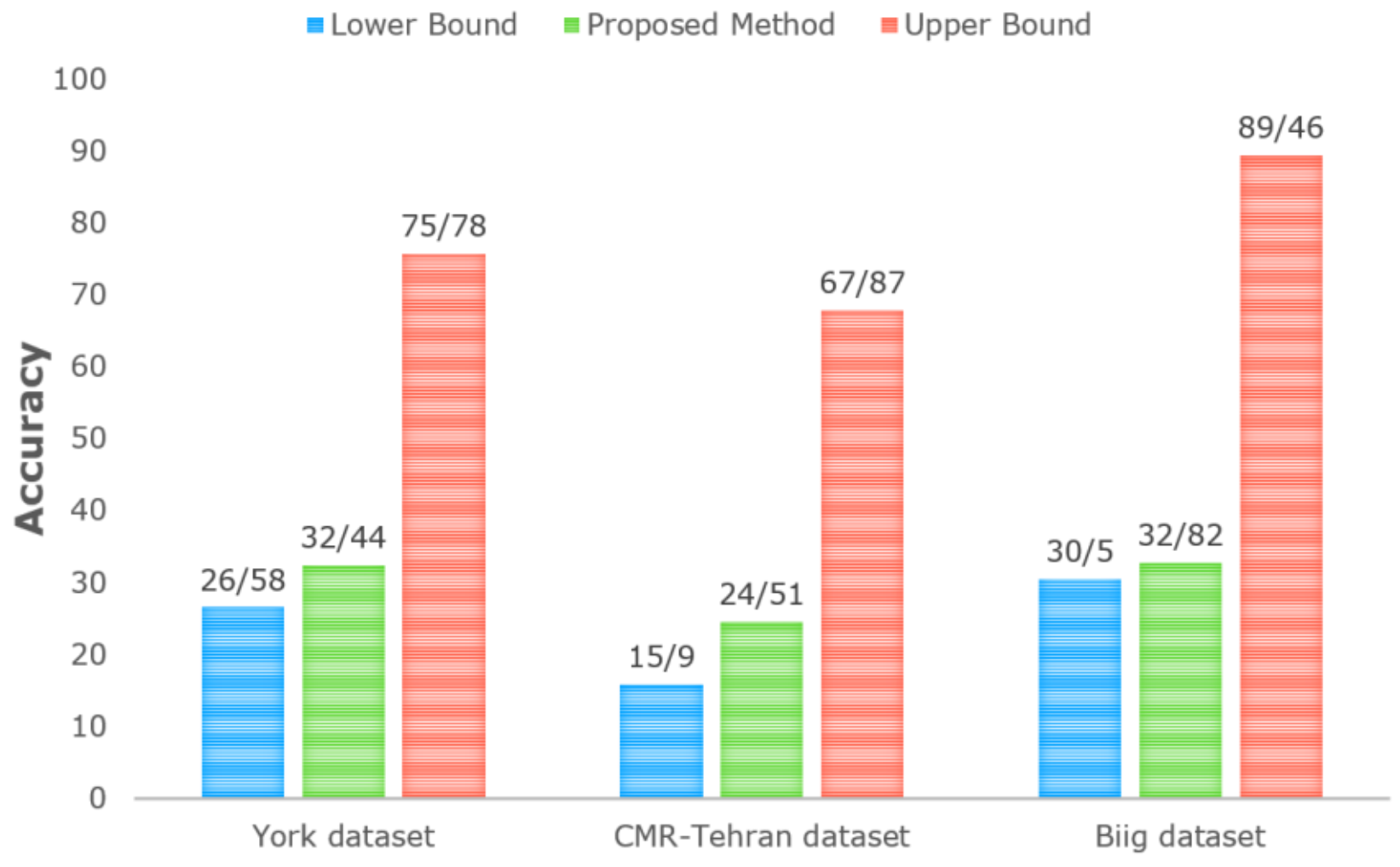

**(a)**

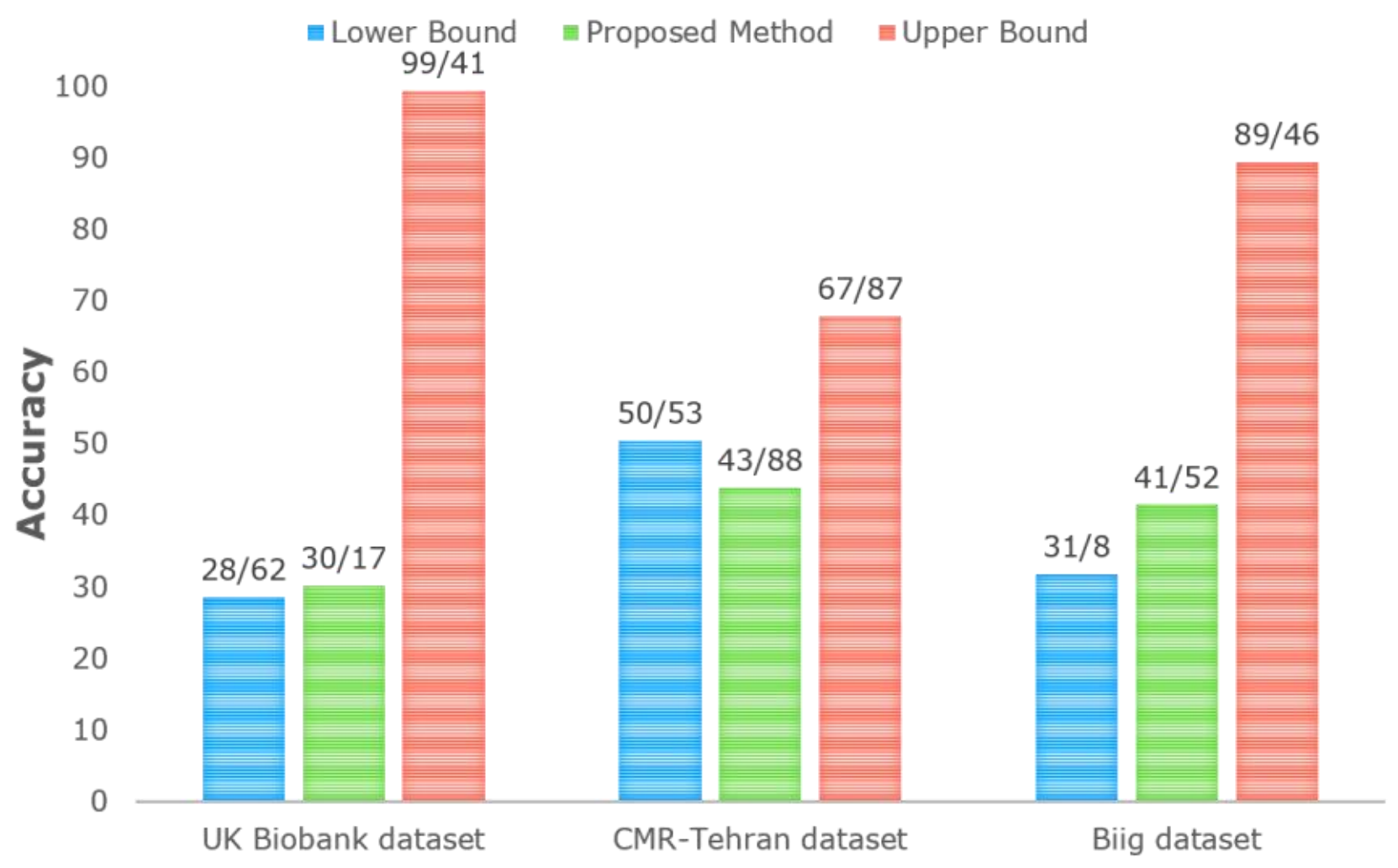

**(b)**

**Figure 7. The proposed domain adaptation results are compared when the source set is (a) the UK Biobank dataset and (b) YU dataset. The lower bound refers to when the training is performed only on the source set, and the upper bound is related to when the training is conducted on the target set.**

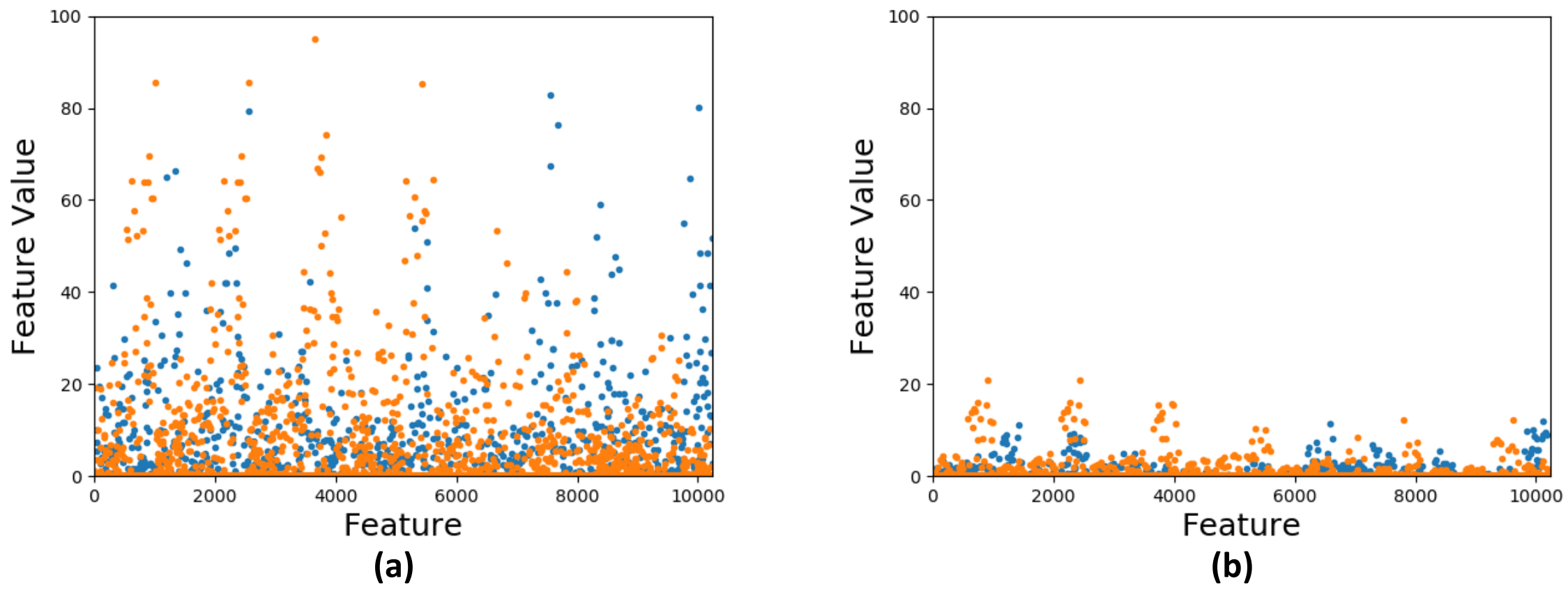

**Figure 8. The effect of domain adaptation on changing the distribution of feature values extracted for 20 images. The feature vectors extracted by the first part of the proposed model in the spatial domain are plotted (a) before and (b) after domain adaptation.**

### 5-3- Spatial versus frequency domain

The time required for supervised training and testing the models in the spatial and frequency domains is compared in Table 6. These times are measured in seconds. The model in the frequency domain can be dramatically learned and tested more quickly. However, this increase in training and testing speed comes with decreased accuracy in this domain. Therefore, we investigated increasing the number of training samples in another experiment to improve the frequency domain's accuracy. The results show that increasing training samples can improve accuracy while training time increases slightly. The results are shown in Table 7 and Figure 9.

**Table 6. Comparison of supervised training and test time differences in proposed deep learning models in spatial and frequency domains for each dataset. The times mentioned are measured in seconds ($Mean \pm std$) based on 4-fold cross-validation. The images related to the UK Biobank, YU, CMR-Tehran, and UCIII datasets are 255775, 106026, 17901 and 4784, respectively. Speed up ratio expresses the increase in training speed in the frequency domain model compared to the spatial domain model.**

| Dataset \ Domain | | Spatial (sec) | Frequency (sec) | Speed up ratio |
|---|---|---|---|---|
| **UK Biobank** | Train | $2249.44 \pm 3.30$ | $258.50 \pm 1.93$ | 8.70 |
| | Test | $32.56 \pm 0.26$ | $7.44 \pm 2.22$ | 4.38 |
| **YU** | Train | $1343.90 \pm 0.83$ | $156.39 \pm 0.17$ | 8.59 |
| | Test | $10.06 \pm 0.28$ | $2.69 \pm 0.91$ | 3.74 |
| **CMR-Tehran** | Train | $277.71 \pm 0.87$ | $52.26 \pm 0.50$ | 5.31 |
| | Test | $2.91 \pm 0.19$ | $0.33 \pm 0.03$ | 8.82 |
| **UCIII** | Train | $133.38 \pm 1.04$ | $42.10 \pm 0.52$ | 3.17 |
| | Test | $1.79 \pm 0.06$ | $0.09 \pm 0.01$ | 19.89 |

**Table 7. Comparison of accuracy based on the different number of training samples in spatial and frequency domains and its effect on training time.**

| | # of training samples | Accuracy | Time (Sec) |
|---|---|---|---|
| **Spatial** | 30,125 | 89.56% | 394.80 |
| **Frequency (Run 1)** | 30,125 | 76.24% | 43.22 |
| **Frequency (Run 2)** | 180,750 | 87.99% | 255.04 |

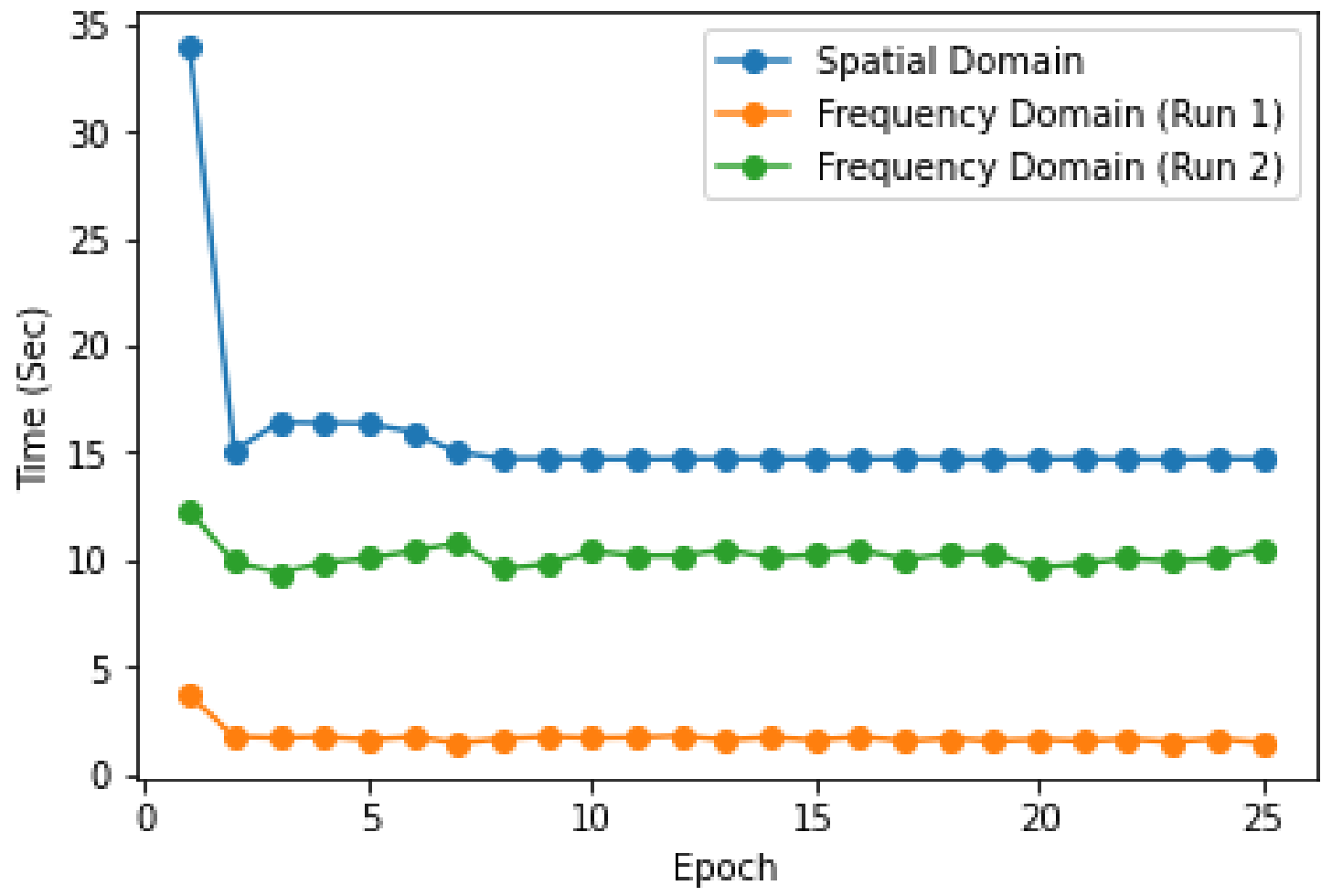

**Figure 9. Training time chart in spatial and frequency domains. In the frequency domain (run 2), 180750 images are used for training. Its accuracy is equal to when using the spatial domain model with 30125 training images. With the increase of training samples, the accuracy in the frequency domain model has become similar to the model's accuracy in the spatial domain. The speed of the training process still varies considerably.**

## 6- Discussion

In this study, we aim to address CMR image quality assessment. For this purpose, we tried to propose a comprehensive model for identifying and reporting common artefacts of these images. Respiratory motion, cardiac motion, Gibbs ringing and aliasing artefacts were examined in this study. Input images with different views were used to provide a model for recognising these artefacts. In addition, the detection of artefacts in spatial and frequency domains was investigated to assess the quality using k-space and with no image reconstruction. Therefore, deep learning models were proposed in both spatial and frequency domains. A deep CNN architecture in the spatial domain and a deep Fourier-based CNN architecture in the frequency domain was used.

One of the main challenges and limitations in deep network training is accessing large amounts of labelled data. Also, while being trained on a dataset and showing promising results, many deep learning approaches may not demonstrate exemplary performance in a new dataset. The concept of domain adaptation was considered to overcome these challenges. Therefore, in this study, several datasets were used. Due to the high quality of UK Biobank images, an attempt was made to select other datasets in terms of distribution, imaging parameters, subjects being imaged, and image quality. Methods were also proposed to add artefacts to CMR images by manipulating the k-space, adding synthetic but realistic artefacts.

Several experiments were designed and performed to evaluate the proposed models. In the first series of experiments, the models were first trained on the prepared datasets supervised. These experiments' results show that the proposed deep learning models can detect the mentioned CMR artefacts. The results show a high AUC for the proposed models, which indicates the suitability of the models to assess the quality of CMR images. High AUC means a higher TP rate can be achieved with a lower FP rate. Besides the high AUC, the precision and recall are also high together, which means the output of the proposed models can be reliable. Also, the AUCs of the spatial and frequency domain models are very close, indicating these models' performance is generally the same.

In this study, we tried to select and augment datasets with considerable distribution gaps. These distribution gaps between datasets are also clearly confirmed by human observers. Despite the significant differences in the available data, the proposed method has improved classification accuracy using domain adaptation. This improvement in results was achieved when no labels were available for the target datasets. Therefore, this approach can partially solve the challenge of data labelling. This issue shows that research in this field can be promising. In one experiment, domain adaptation did not show the desired result, reported in existing studies in this field (Ganin and Lempitsky 2015, Gholami, Sahu et al. 2020).

Another achieved goal of this study is to increase the speed of deep learning-based models. For this purpose, Fourier-based CNNs was used to propose the model. Based on the analysis of the results, these networks can perform well while increasing training and testing speed. Using the model in the frequency domain eliminates the need to define fixed-size kernels in convolution operations. Defining kernels with the same size as the input image means that the extracted features are not just local. There is also no need to move the sliding window over the entire image, which is time-consuming. In addition, large size images can also be used in these models because the complexity of feature extraction operations is much less than the spatial domain. According to the experiments performed in the results Section, the desired accuracy can be achieved in a shorter time by increasing the number of samples.

The proposed model in the frequency domain can create the ability to directly evaluate the k-space quickly after filling in terms of quality. Besides shortening the image quality assessment time compared to when spatial domain models are used, the cost of image reconstruction using fast Fourier transform, which is $O(n \log n)$ for each k-space, is avoided. Also, the frequency domain model can be used immediately after filling any k-space, and if the quality is not satisfactory, image acquisition can be repeated without wasting time.

CMR image quality control involves examining the full heart coverage, signal-to-noise ratio, and imaging artefacts. A different approach is taken to explore each of these cases. In some studies (Zhang, Gooya et al. 2017, Zhang, Gooya et al. 2018, Zhang, Pereañez et al. 2018), whole heart or left ventricle coverage were examined. In these studies, short-axis cine CMR volumes are investigated for the presence or absence of apex and basal slices. Other researches (Tarroni, Oktay et al. 2018, Oksuz, Ruijsink et al. 2019, Tarroni, Bai et al. 2020) have focused on CMR imaging artefacts. The proposed models of these studies are 3D models based on the evaluation of artefacts in a 3D CMR volume. Since physicians ultimately examine CMR images in a 2D view, the current study used a two-dimensional model to identify the artefacts. Besides, in this study, four types of the most common CMR artefacts were evaluated, which is superior to the previous studies regarding the number of artefacts examined.

Furthermore, due to the limited access to data labels in medical tasks, a domain adaptation method was proposed to overcome this problem. This issue has not been considered in the previous studies of the field, and we tried to address it in this study. Investigation of reducing training and quality control time using the FCNN model to identify artefacts more quickly is another contribution of this study that is not seen in the related studies.

## 7- Conclusion

In this study, the CMR image quality assessment was examined. Two models, one in the spatial domain and one in the frequency domain, were proposed. These models can identify four common artefacts in CMR images, including respiratory motion, cardiac motion, Gibbs ringing and aliasing. Besides presenting models to evaluate the quality of these images, several other objectives were considered. Suggesting the methods to add artefacts to images and generate synthetic but realistic degraded CMR images, assessing the increase in training and testing time in the frequency domain model, using the model to evaluate quality using k-space before image reconstruction, using domain adaptation to overcome the challenge of not having access to data labels, are another characteristics of this study. The results show well the ability of the proposed models to meet the above objectives. Presenting more comprehensive models covering more artefacts and using new methods such as meta-learning will be among the purposes of the future study.

## Acknowledgements

This research has been done using the UK Biobank database under Application 11350.